\documentclass[paper]{ieice}
\usepackage{graphicx,xcolor}
\usepackage[fleqn]{amsmath}
\usepackage{multirow}
\usepackage[psamsfonts]{amssymb}
\usepackage{amsxtra}
\usepackage{amsmath}
\usepackage{ulem}
\usepackage{soul}
\usepackage{threeparttable,comment}
\usepackage{cite,array}
\usepackage{mathrsfs,color}
\usepackage{mathtools}

\usepackage{threeparttable,bm,amsmath,graphicx,amssymb,comment,array,multirow,threeparttable,cite,calligra,color}
\usepackage[subrefformat=parens]{subcaption}
\usepackage{comment}
\usepackage{amssymb}
\usepackage{amsfonts}
\usepackage{color}

\usepackage{cite}
\usepackage{textcomp}
\usepackage{amsmath,graphicx}

\makeatletter
\newcommand{\figcaption}[1]{\def\@captype{figure}\caption{#1}}
\newcommand{\tblcaption}[1]{\def\@captype{table}\caption{#1}}


\makeatletter
\renewcommand\subsubsection{\@startsection{subsubsection}{3}{\z@}%
{0.5ex\@plus 0ex \@minus -.5ex}%
{0.5ex\@plus 0ex}
{\normalfont\normalsize\itshape}
}
\makeatother

\field{}
\title{A Privacy-Preserving Machine Learning Scheme Using EtC Images } \authorlist{
\authorentry[kawamura-ayana@ed.tmu.ac.jp]{Ayana KAWAMURA}{n}{tmu}\MembershipNumber{}
\authorentry[kinoshita-yuma@ed.tmu.ac.jp]{Yuma KINOSHITA}{s}{tmu}\MembershipNumber{}
\authorentry[nakachi.takayuki@lab.ntt.co.jp]{Takayuki NAKACHI}{m}{ntt}\MembershipNumber{}
\authorentry[sayaka@tmu.ac.jp]{Sayaka SHIOTA}{m}{tmu}\MembershipNumber{}
\authorentry[kiya@tmu.ac.jp]{Hitoshi KIYA}{f}{tmu}\MembershipNumber{}
}
\affiliate[tmu]{The authors are with Tokyo Metropolitan University, Hino-shi, 191-0065 Japan}
\affiliate[ntt]{The authors are with NTT Network Innovation Laboratories, Kanagawa, 239-0847 Japan}
\received{2015}{1}{1}
\revised{2015}{1}{1}
\begin{document}
\newcommand{\red}[1]{\textcolor{black}{#1}}
\maketitle
\begin{summary}
We propose a privacy-preserving machine learning scheme with encryption-then-compression (EtC) images, where EtC images are images encrypted by using a block-based encryption method proposed for EtC systems with JPEG compression. 
In this paper, a novel property of EtC images is first discussed, although EtC ones was already shown to be compressible as a property. 
The novel property allows us to directly apply EtC images to machine learning algorithms non-specialized for computing encrypted
data. 
In addition, the proposed scheme is demonstrated to provide no degradation in the performance of some typical machine learning algorithms including the support vector machine algorithm with kernel trick and random forests under the use of z-score normalization. 
A number of facial recognition experiments with are carried out to confirm the effectiveness of the proposed scheme.
\end{summary}

\vspace{-0.1in}
\section{Introduction}
\label{sec:intro}

Cloud computing and edge computing have been spreading in many fields with the development of cloud services. 
However, cloud environments have serious issues for end users, such as the unauthorized use of services, data leaks, and privacy being compromised due to unreliable providers and some accidents \cite{huang2014survey}. 
Because of such a situation, various methods have been proposed for privacy-preserving computing in cloud environments. 
In this paper, we propose a privacy-preserving machine learning scheme using compressible encrypted images, called
Encryption-then-Compression (EtC) images. 
Machine learning requires a huge amount of data for training a model, and moreover most of data include sensitive personal information.

One of ways for privacy-preserving computing is to use a perceptual image encryption method which aims to protect visual information on plain ones. 
Compared to information theory- based-encryption\cite{Shokri2015,Phong2018,Phong2018corr,Dowlin2016,Wang2018,Yang2017,Wang2016,Saxe2018} such as multi-party computation and homomorphic encryption, perceptual encryption methods have a number of advantages. 
Use of perceptual encryption allows us to directly apply machine learning algorithms without increasing computational costs. 
In other words, there is no need to prepare algorithms specialized for computing encrypted data.
Therefore, privacy-preserving computing schemes with visually protected images have been studying for secure cloud services and computing.

Some of them can produce compressible encrypted images\cite{Chuman2018,Warit2019APSIPA,Itier2019,watanabe2015encryption}, but they have never considered applying encrypted images to machine learning algorithms. 
In contrast, a number of methods can be applied to machine learning algorithms\cite{Nakamura2015,Nakamura2016,maekawa2019,warit2019ICIP,warit2019Access}, but the encrypted images are not compressible. 
To reduce the amount of data, images are required to be compressed in the transmission and the storage in general.

Consequently,  we focus on EtC images as compressible ones. 
EtC images are images encrypted by using a block-based encryption method proposed for EtC systems with JPEG compression \cite{Chuman2018,Warit2019APSIPA,Kuri_2017}. 
So far, the safety of the EtC systems has been evaluated on the basis of the key space under the assumption of brute-force attacks, and robustness against jigsaw puzzle attacks has been discussed\cite{CHUMAN2017ICASSP,CHUMAN2017ICME}. 
In this paper, in addition to compressible images, EtC images is shown to have a novel property, i.e. learnable images, under the use of z-score normalization.
The novel property allows us to securely compute typical machine learning algorithms such as support vector machine (SVM) and random forests without any degradation in performances. 
In an experiment, the proposed scheme is applied to a facial recognition algorithm with classifiers to confirm the effectiveness of the scheme under the use of SVM and random forests.

\vspace{-0.1in}
\section{Preparation}
\label{sec:pre}
\subsection{EtC image}
\begin{figure}[t]
\begin{center}
\includegraphics[width=8cm]{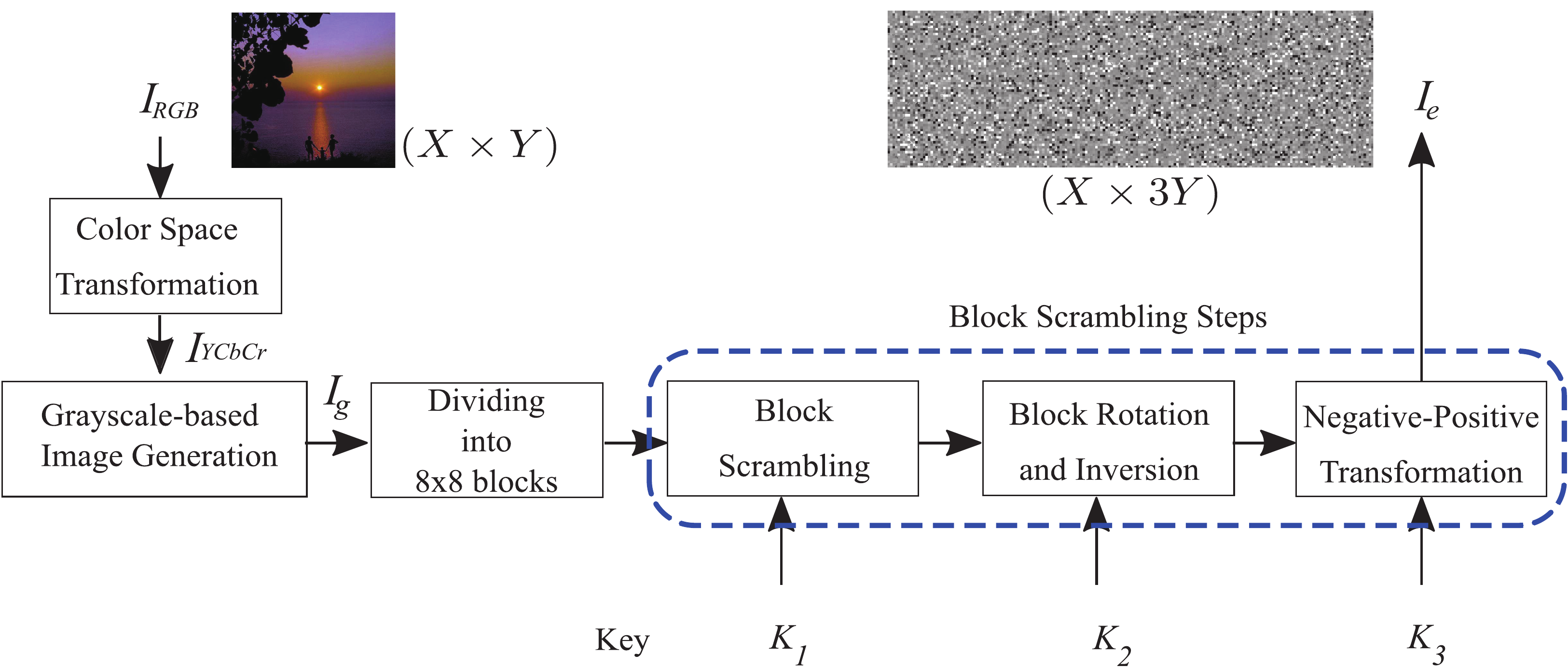}
\caption{EtC image \cite{Chuman2018,Warit2019APSIPA}}
\label{enc}
\end{center}
\end{figure}

In this paper, EtC images will be shown to have a novel property, and the novel property allows us to carry out privacy-preserving machine learning.
EtC images are images encrypted by the encryption method\cite{Chuman2018,Warit2019APSIPA}, which was proposed for Encryption-then-Compression systems with JPEG compression.
EtC images are robust enough against various ciphertext-only attacks, including jigsaw puzzle solver attacks\cite{CHUMAN2017ICASSP,CHUMAN2017ICME}.

The procedure of generating EtC images is summarized here (See Fig.\ref{enc}).

\noindent 1) \red{Transform a} color image in the RGB color space ($I_{RGB}$) with $X\times Y$ pixels into an image ($I_{YCbCr}$) in the YCbCr color space, \textcolor{black}{ where the color space transformation is carried out on the basis of the transformation in \cite{color_trans}.}

\noindent 2) Split $I_{YCbCr}$ into \red{three} channels \red{that} can be represented \red{as} $i_Y$, $i_{Cb}$, \red{and} $i_{Cr}$.
Then\red{,} integrate the three channels for generating a grayscale-based image ($I_g$) \textcolor{black}{with $X \times (3 \times Y)$} pixels \textcolor{black}{(See Fig.\ref{enc})}.

\noindent 3) Divide $I_g$ into blocks with $B_x$×$B_y$ pixels, and permute randomly the divided blocks \red{by} using a random integer generated by a secret key $K_1$.

\noindent 4) Rotate and invert randomly each block \red{by} using a random integer generated by a key $K_2$.

\noindent 5) Apply negative-positive transformation to each block \red{by} using a random binary integer generated by a key $K_3$.
In this step, a transformed pixel value in \red{the} $i$th block $B_i$, $p'$ is computed by
\begin{equation}
\begin{cases}
p'=p    & (r(i)=0) \\
p'=255-p & (r(i)=1)
\end{cases},
\label{negaposi}
\end{equation}
where $r(i)$ is a random binary integer generated by $K_3$ under the probability $P(r(i))=0.5$\red{,} and $p$ is the pixel value of an original image with 8 bpp.

 In this paper, images encrypted \red{with} the above steps are referred to as EtC images.
Steps 1 and 2 are \red{skipped} when grayscale images are encrypted.

\subsection{Scenario}
\begin{figure}[t]
\begin{center}
\includegraphics[width=7.5cm]{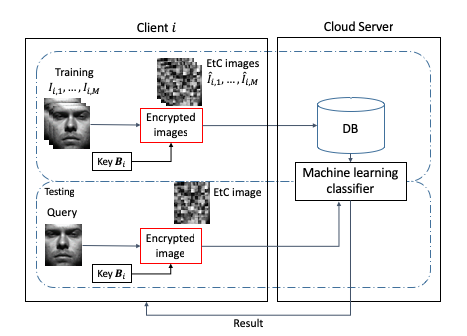}
\caption{Privacy-preserving machine learning}
\label{svmsystem}
\end{center}
\end{figure}

 Figure \ref{svmsystem} illustrates the scenario used in this paper. \textcolor{black}{EtC images aim to protect visual information that allow us to identify an individual, the time and the location of the taken photograph as well as most other perceptual image encryption methods.} 
In the training \red{task}, client $i,i=1,...,N$ prepares training images $I_{i,j},j=1,...,M$.
Next\red{,} the client creates encrypted images $\hat{I}_{i,j}$ by \red{using} a secret key set $\bm{B}_i=\{K_{i_1},K_{i_2},K_{i_3}\}$ and send\red{s} the encrypted ones to a cloud server, where the dimensionality of the encrypted images may be \red{reduced} in order to avoid the effects of the curse of dimensionality\cite{Boufounos2016}.
Finally, the cloud server carries out learning with the encrypted data for a classification problem.

 In the testing \red{task}, client $i$ creates an encrypted image with $\bm{B}_i$ as a query and sends the image to the server.
The server carries out a classification problem with a model prepared in advance. Finally, the server returns a result to client $i$.

Cloud servers are not trusted in general, so there are some security issues with cloud environments such as the unauthorized use of services, data leaks, and privacy being compromised.
Because of such a situation, we propose a privacy-preserving machine learning scheme in this paper.
In this scenario, cloud servers are not given secret keys and cannot obtain any visual information.

\section{Proposed privacy-preserving machine learning}
\label{sec:pro}
In this paper, we propose a privacy-preserving machine learning scheme using EtC images.
A novel property of EtC images is shown here, and the property is applied to privacy-preserving machine learning.

\subsection{Novel property of EtC images}

Let us transform a grayscale image $I_{i,j}$ with $X\times Y$ pixels into a vector $\bm{T}_{i,j}=\{p_{i,j}(0),\ldots,p_{i,j}(d-1)\}^\mathrm{T}\in\mathbb{R}^d, d=X\times Y$, where $p_{i,j}(k),k=0,1,\ldots,d-1$ is a pixel value of $I_{i,j}$.


\vspace{2ex}
\noindent {\textcolor{black}{A. Block scrambling, block rotation and inversion}}

 As shown in Fig.\ref{enc}, block scrambling, and block rotation and inversion \red{are carried out for} permut\red{ing} pixels.
 Those operations are easily shown to be represented as a permutation matrix.
For example\red{,} a permutation matrix $\bm{Q}_{i}$ is give\red{n} as, for $d=3$\red{,}
\begin{equation}
\bm{Q}_{i}=\left(
\begin{array}{ccc}
1&0&0\\
0&0&1\\
0&1&0\\
\end{array}
\right),
\end{equation}
where $\bm{Q}_{i}$ has only one element of 1 in each row or each column\red{,} and others are 0, so $\bm{Q}_{i}$ becomes an orthogonal matrix.
The orthogonal matrix meets the equation
\begin{equation}
\bm{Q}_{i}^\mathrm{T}\bm{Q}_{i}=\bm{E},
\label{Q}
\end{equation}
where $\bm{E}$ is an identity matrix, and $\mathrm{T}$ indicates transpose.

 A encrypted vector $\hat{\bm{T}}_{i,j}$ is computed by using $\bm{Q}_{i}$ as
\begin{equation}
\hat{\bm{T}}_{i,j}=\bm{Q}_{i}\bm{T}_{i,j}.
\label{QBI}
\end{equation}
Therefore, $\hat{\bm{T}}_{i,j}=\{\hat{p}_{i,j}(0),\ldots,\hat{p}_{i,j}(d-1)\}^\mathrm{T}$ in Eq.(\ref{QBI}) meets the properties in Eqs.(\ref{kyori}) and (\ref{naiseki}), due to the orthogonality of $\bm{Q}_{i}$\cite{Nakamura2016}, where $\hat{p}_{i,j}(k)$ corresponds to a pixel value of an EtC image generated under the use of block scrambling, and block rotation and inversion operations.

\noindent Property1: Conservation of Euclidean distances
\begin{equation}
\|\bm{T}_{i,j}-\bm{T}_{s,t}\|^2=\|\hat{\bm{T}}_{i,j}-\hat{\bm{T}}_{s,t}\|^2.
\label{kyori}
\end{equation}

\noindent Property2: Conservation of inner products
\begin{equation}
\langle\bm{T}_{i,j},\bm{T}_{s,t}\rangle=\langle\hat{\bm{T}}_{i,j},\hat{\bm{T}}_{s,t}\rangle.
\label{naiseki}
\end{equation}
$\bm{T}_{s,t}$ is a transformed vector from image $I_{s,t}$ where $I_{s,t}$, $t\in \lbrace 1,2,\ldots,M \rbrace $ is an image of client $s, s\in \lbrace 1,2,\ldots,N \rbrace$.

\vspace{2ex}
\noindent {\textcolor{black}{B. Negative-positive transformation}}

 Next, let us consider the influence of negative-positive transformation.
In the case of using the transformation in Eq.(\ref{negaposi}), the relation between a pixel value $p'_{i,j}(k)=255-p_{i,j}(k)$ and another $p'_{s,t}(k)=255-p_{s,t}(k)$ is given by
\begin{equation}
\|p'_{i,j}(k)-p'_{s,t}(k)\|^2=\|p_{i,j}(k)-p_{s,t}(k)\|^2.
\end{equation}
This relation shows that the Euclidean distance between $p_{i,j}(k)$ and $p_{s,t}(k)$ is preserved.
However, since the relation 
\begin{eqnarray}
p'_{i,j}(k)\cdot p'_{s,t}(k) &=& (255-p_{i,j}(k))\cdot (255-p_{s,t}(k)) \nonumber \\
&\neq& p_{i,j}(k)\cdot p_{s,t}(k) ,
\end{eqnarray}
the inner product is not preserved.
Consequently, the negative-positive transformation operation preserves only Euclidean distance between $p_{i,j}(k)$ and $p_{s,t}(k)$.

\vspace{2ex}
\noindent {\textcolor{black}{C. Negative-positive transformation with z-score normalization}}

Next\red{,} we consider using z-score normalization\cite{Jain2005}, which is a well-known data normalization method for machine learning.
In z-score normalization, a value $p_{i,j}(k)$ is replaced with $z_i$ like
\begin{equation}
z_{i,j}(k)=(p_{i,j}(k)-\bar{P})/S,
\label{normalization}
\end{equation}
where $\bar{P}$ is the mean value of data\red{,} and $S$ is the standard deviation given by
\begin{equation}
S=\sqrt{\frac{\sum_{i=1}^{N}(p_{i,j}(k)-\bar{P})^2}{N}}.
\end{equation}
 Therefore, in negative-positive transformation, Eq.(\ref{normalization}) is given as
\begin{eqnarray}
z'_{i,j}(k)&=&-\frac{p'_{i,j}(k)-\bar{P'_k}}{S'} \nonumber \\
&=&\frac{(255-p_{i,j}(k))-(255-\bar{P_k})}{S'} \nonumber \\
&=&-\frac{p_{i,j}(k)-\bar{P_k}}{S}=-z_{i,j}(k),
\label{znp}
\end{eqnarray}
where
\begin{equation}
\bar{P_k}=\frac{1}{N\times M}\sum_{i=1}^{N}\sum_{j=1}^{M}p_{i,j}(k)
\end{equation}
\begin{equation}
\bar{P'_k}=\frac{1}{N\times M}\sum_{i=1}^{N}\sum_{j=1}^{M}p'_{i,j}(k) = 255 - \bar{P_k}
\end{equation}
\begin{eqnarray}
S'&=&\sqrt{\frac{\sum_{i=1}^{N}\sum_{j=1}^{M}((255-p_{i,j}(k))-(255-\bar{P_k}))^2}{N\times M}} \nonumber \\
&=&\sqrt{\frac{\sum_{i=1}^{N}\sum_{j=1}^{M}(-p_{i,j}(k)+\bar{P_k})^2}{N\times M}}=S.
\end{eqnarray}
Eq.(\ref{znp}) means that the normalized value $z'_{i,j}(k)$ of $p'_{i,j}(k)$ becomes the sign inverted value of the normalized value $z_{i,j}(k)$ of $p_{i,j}(k)$.
A sign inversion matrix can be expressed as an orthogonal matrix, so the inner product is preserved under the use of z-score normalization.

Hence, in the case of applying z-score normalization to EtC images,  negative-positive transformation allows us to maintain the inner products.
As a result, EtC images can maintain not only the Euclidean distance but also the inner products under the use of z-score normalization.




\subsection{Privacy-preserving machine learning}
\noindent{\textcolor{black}{A. SVM with kernel trick}}

 We first focus on SVM as an example of machine learning algorithms.
In SVM computing, we input a feature vector $\bm{x}$ to the discriminant function as
\begin{equation}
\label{svm}
f(x)={\rm sign}(\bm{\omega}^\mathrm{T}\bm{x}+b)
\end{equation}
with
\begin{equation}
{\rm sign}(u)=
\begin{cases}
1    & (u>1) \nonumber \\
-1 & (u\le0) \nonumber
\end{cases},
\end{equation}
where $\bm{\omega}$ is a weight parameter vector, \red{and} $b$ is a bias.

SVM has also a technique called \red{"}kernel trick\red{"}.
When the kernel trick is applied to Eq.(\ref{svm}), the equation is given by
\begin{equation}
f(x)={\rm sign}(\bm{\omega}^\mathrm{T}\phi(\bm{x})+b).
\end{equation}
The function $\phi(\bm{x}):\mathbb{R}^d\longrightarrow\mathcal{F}$ maps an input vector $\bm{x}$ on high dimensional feature space $\mathcal{F}$, where $d$ is the number of the dimensions of features.
The kernel function of two vectors $\bm{x}_i$, $\bm{x}_j$ is defined as
\begin{equation}
K(\bm{x}_i,\bm{x}_j)=\langle\phi(\bm{x}_i),\phi(\bm{x}_j)\rangle,
\end{equation}
where $\langle\cdot,\cdot\rangle$ is an inner product.

Typical kernel functions such as the radial basis function (RBF) kernel, linear one and polynomial one are based on the distance or the inner products between two vectors.
For example, RBF kernel is based on the Euclidean distance and polynomial kernel is based on the inner products, as
\begin{equation}
K(\bm{x}_i,\bm{x}_j)={\rm exp}(-\gamma\|\bm{x}_i-\bm{x}_j\|^2)
\label{rbfkernel}
\end{equation}
\begin{equation}
K(\bm{x}_i,\bm{x}_j)=(1+\langle\bm{x}_i,\bm{x}_j\rangle)^l,
\end{equation}
where $\gamma$ is a hyperparameter \red{for} decid\red{ing} the complexity of boundary determination, \red{and} $l$ is a parameter \red{for} decid\red{ing} the degree of the polynomial.

 In the case of using \red{the} RBF kernel, the following relation is satisfied from property 1 and Eq.(\ref{rbfkernel})\red{:}
\begin{eqnarray}
K(\hat{\bm{T}}_{i,j},\hat{\bm{T}}_{s,t})&=&{\rm exp}(-\gamma\|\hat{\bm{T}}_{i,j}-\hat{\bm{T}}_{s,t}\|^2) \nonumber \\
&=&K(\bm{T}_{i,j},\bm{T}_{s,t}).
\label{eqrbfsame}
\end{eqnarray}
Eq. (\ref{eqrbfsame}) meets that EtC images do not \red{have} any influence on the kernel function.
Therefore, when using kernel functions based on the Euclidean distance\red{,} EtC images provide the same result as that of plain images.

 In addition, from property 2, we can also use a kernel based on the inner products between two vectors under the use of z-score normalization.
\red{The p}olynomial kernel and linear kernel are in this class.

 Next, we consider binary classification.
A dual problem \red{for} implement\red{ing} a SVM classifier with encrypted images is expressed as

\footnotesize
\begin{align}
&\max_\alpha\left(-\frac{1}{2}\sum_{\substack{i,s\in N \\ j,t\in M}}^{} \alpha_{i,j}\alpha_{s,t}y_{i,j}y_{s,t}\langle\phi(\hat{\bm{T}}_{i,j}),\phi(\hat{\bm{T}}_{s,t})\rangle+\sum_{\substack{i,s\in N \\ j,t\in M}}^{}\alpha_{i,j}\right) \nonumber \\
& s.t. \sum_{\substack{i,s\in N \\ j,t\in M}}^{}\alpha_{i,j}y_{i,j}=0,\;\;0\le \alpha_{i,j}\le C,
\end{align}
\normalsize
where $y_{i,j}$ and $y_{s,t}\in\{+1,-1\}$ are correct labels for each \red{piece of} training data, $\alpha_{i,j}$ and $\alpha_{s,t}$ are dual variables and $C$ is a regular coefficient.
If we use a kernel function described above, the inner product $K(\hat{\bm{T}}_{i,j},\hat{\bm{T}}_{s,t})=\langle\phi(\hat{\bm{T}}_{i,j}),\phi(\hat{\bm{T}}_{s,t})\rangle$ is equal to $K(\bm{T}_{i,j},\bm{T}_{s,t})$.
Therefore, even in the case of using encrypted images, the dual problem with encrypted images is reduced to the same problem as that of the plain images.
This conclusion means that the use of the encrypted ones \red{has} no influence \red{on} the performance of the SVM classifier.

\vspace{2ex}
\noindent {\textcolor{black}{B. Other machine learning algorithms}}

In addition to SVM, the proposed method can be applied to other machine learning algorithms based on the Euclidean distance or the inner products.
For example, EtC images can be applied to the k-nearest neighbor (kNN) algorithm and random forests as well.
The kNN algorithm is based on the Euclidean distance between training data $\bm{x}_i$ and testing data $\bm{x}$ like $\|\bm{x}_i-\bm{x}\|^2$, and the output is a class membership.
An object is classified by a plurality vote of its neighbors with the object being assigned to the class most common among its $k$ nearest neighbors, where $k$ is a positive integer.
Therefore, we obtain the same result as plain images under the use of the kNN algorithm, as well as for SVM.

We can also apply the proposed method to random forests. 
Random forests\cite{Breiman2001} offers results determined by the relative relationship of feature vector elements among images.
When the relative relationship between two images is not changed, or all relation are inverted, random forests can provide the same result as plain images. 
EtC images have this property under key condition 1, so we can obtain the same result as plain images even under the use of random forests.

\subsection{Relation among keys}
 As shown in Fig.\ref{svmsystem}, an encrypted image $\hat{I}_{i,j}$ is generated from $I_{i,j}$ by using a key set $\bm{B}_i$.
Two relations among keys are summarized, here.

\vspace{2ex}
\noindent{\textcolor{black}{A. Key condition 1 ($\bm{B}_1=\bm{B}_2=\ldots=\bm{B}_N$)}}

 The first key choice is to use a common key \red{for} all clients, namely, $\bm{B}_1=\bm{B}_2=\ldots=\bm{B}_N$.
In this case, all encrypted images satisfy the properties described in 3.1 and 3.2, so the SVM classifier has the same performance as that of using the original images.

\vspace{2ex}
\noindent{\textcolor{black}{B. Key condition 2 ($\bm{B}_1\neq \bm{B}_2\neq\ldots\neq \bm{B}_N$)}}

 The second key choice is to use a different key \red{for} each client, namely\red{,} $\bm{B}_1\neq \bm{B}_2\neq\ldots\neq \bm{B}_N$.
In this case, the properties are satisfied only among images with a common key.
This key condition allows us to enhance the robustness of security against various attacks as discussed later.

\section{Experiment and Discussion}
\label{sec:sim}
 The proposed scheme was applied to fac\red{ial} recognition experiments \red{that} were carried out with SVM and random forests.

\subsection{Experimental setup}
\begin{figure}[t]
\centering
  \begin{tabular}{c c c c}
     \begin{minipage}[b]{0.15\hsize}
 	     \centering\includegraphics[width = 1.5cm]{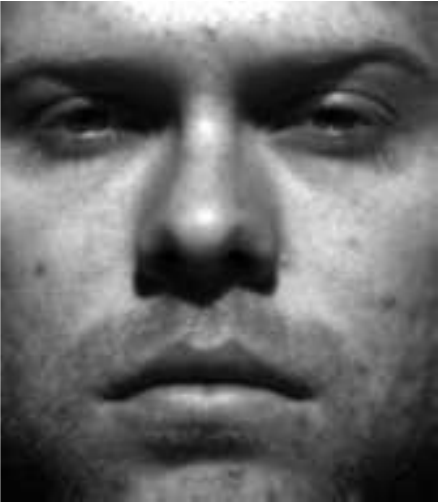}
     \end{minipage}
      &
     \begin{minipage}[b]{0.15\hsize}
        \centering\includegraphics[width = 1.5cm]{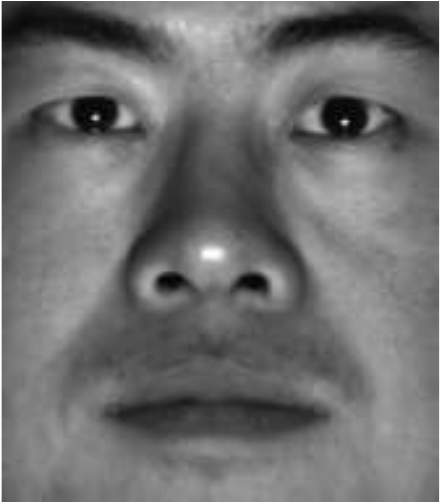}
    \end{minipage}
      &
    \begin{minipage}[b]{0.15\hsize}
   	 \centering\includegraphics[width = 1.4cm]{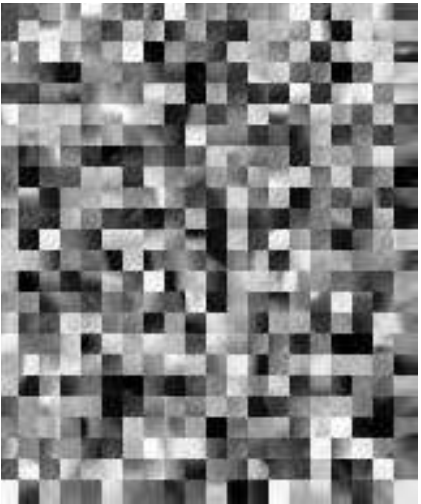}
    \end{minipage}
      &
    \begin{minipage}[b]{0.15\hsize}
      \centering\includegraphics[width = 1.4cm]{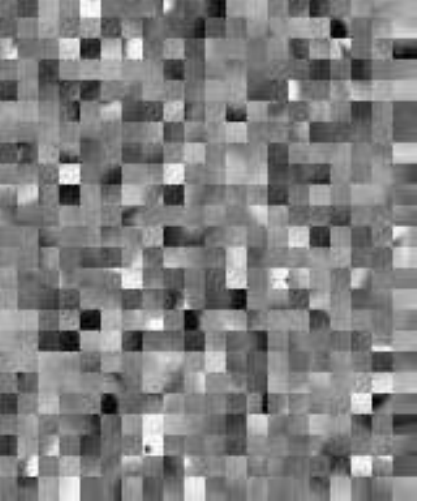}
    \end{minipage}
  \\
    \multicolumn{2}{c}{(a) Original image}
     &
    \multicolumn{2}{c}{(b) EtC image}
  \end{tabular}
\caption{\red{E}xample of facial images}
\label{encrypttemplate}
\end{figure}

 We used \red{the} Extended Yale Face Database B\cite{Geo2001}\red{, which} consists of 38 $\times$ 64 = 2432 frontal facial images with 192 $\times$ 160 pixels \red{for} $N=38$ persons.
$M=64$ images for each person were divided into half randomly for training data samples and queries.
$B_x$×$B_y=8\times 8$ was used. 
The random projection method\cite{Kaski1998} was carried out as a dimensionality reduction method in the encrypted domain.
 When using the random projection method, original $d$-dimensional vectors are projected to a $d_r$-dimensional subspace, ($d_r\leq d$), \red{by} using a random $d_r\times d$ matrix.
 The average and variance of the Euclidean length of each column is respectively 0 and 1.
Dimensionality \red{was reduced} with reduction ratio\red{s} of 1/20, 1/40, 1/60\red{, and} 1/80, so $d=192\times 160=30720$ \red{was} reduced to $d_r=1536$ for \red{a ratio of} 1/20.
After the dimensionality reduction, we applied z-score normalization to the images.
Figure \ref{encrypttemplate} shows examples of original images and the EtC images.

\begin{table}[t]
\begin{center}
\caption{Machine spec}
\label{spec}
\begin{tabular}{|c|c|c|}
\hline
Processor & Intel Core i5-6500 3.20GHz \\ \hline
Memory & 8 GB \\ \hline
OS & Ubuntu 16.04 LTS \\ \hline
Software & MATLAB R2018a \\ \hline
\end{tabular}
\end{center}
\end{table}

\subsection{Effects of z-score normalization}
The effects of z-score normalization were first confirmed by using plain images.
The simulation was run on a PC, with a 3.2 GHz processor and main memory of 8 Gbytes (see Table \ref{spec}).
SVM classifiers with the linear kernel were used in this simulation. 

In fac\red{ial} recognition with SVM classifiers, one classifier is created for each client.
The classifier outputs a predicted class label and classification score for each query image $I_q$.
The classification score is the distance from \red{a} query to the boundary rang\red{e}.
The relation between the classification score $S_q$ and a threshold $\tau$ for \red{a} positive label of $I_q$ is given as
\begin{equation}
if\;S_q\leq\tau\;then\;accept;\;else\;reject.
\label{accept}
\end{equation}

In this experiment, equal error rate (EER), at which false accept rate (FAR) is equal to false reject rate (FRR), were used to evaluate the performance.

\textcolor{black}{Table \ref{linearEER} shows the EER values for the SVM with the linear kernel. The values of EER with z-score normalization are smaller than those without the normalization. This means z-score normalization improves the EER performance under the use the SVM classifier.}
Table \ref{lineartime} shows training time in the same situation as in Table \ref{linearEER}.
By applying z-score normalization, training time was reduced to less than 1/100.
Therefore, z-score normalization allows us not only to improve EER values, but also to reduce training time.
In addition, z-score normalization provides a novel property to EtC images as described in 3.1.C.

\begin{table}[t]
\begin{center}
\caption{{$\scriptstyle \mbox{EER values for SVM with linear kernel}\atop \scriptstyle \mbox{(plain images)}$}}
\label{linearEER}
\begin{tabular}{|c|c|c|}
\hline
\begin{tabular}{c}reduction\\ ratio\end{tabular} & \begin{tabular}{c}without z-score\\ normalization\end{tabular} & \begin{tabular}{c}with z-score\\ normalization\end{tabular} \\ \hline
1/40 & 0.369 & 0.0248 \\ \hline
1/80 & 0.349 & 0.0297 \\ \hline
\end{tabular}
\end{center}
\end{table}

\begin{table}[t]
\begin{center}
\caption{{$\scriptstyle \mbox{Training time for SVM with linear kernel}\atop \scriptstyle \mbox{(plain images)}$}}
\label{lineartime}
\begin{tabular}{|c|c|c|}
\hline
\begin{tabular}{c}reduction\\ ratio\end{tabular} & \begin{tabular}{c}without z-score\\ normalization\end{tabular} & \begin{tabular}{c}with z-score\\ normalization\end{tabular} \\ \hline
1/40 & 5331[sec] & 23.48[sec] \\ \hline
1/80 & 5201[sec] & 16.74[sec] \\ \hline
\end{tabular}
\end{center}
\end{table}

\subsection{Experimental results with SVM}
 In this experiment with SVM, \red{the} RBF kernel and linear kernel were used.

\begin{figure}[t]
 \begin{minipage}[b]{1\linewidth}
  \centering
  \includegraphics[keepaspectratio, width=7.5cm]
  {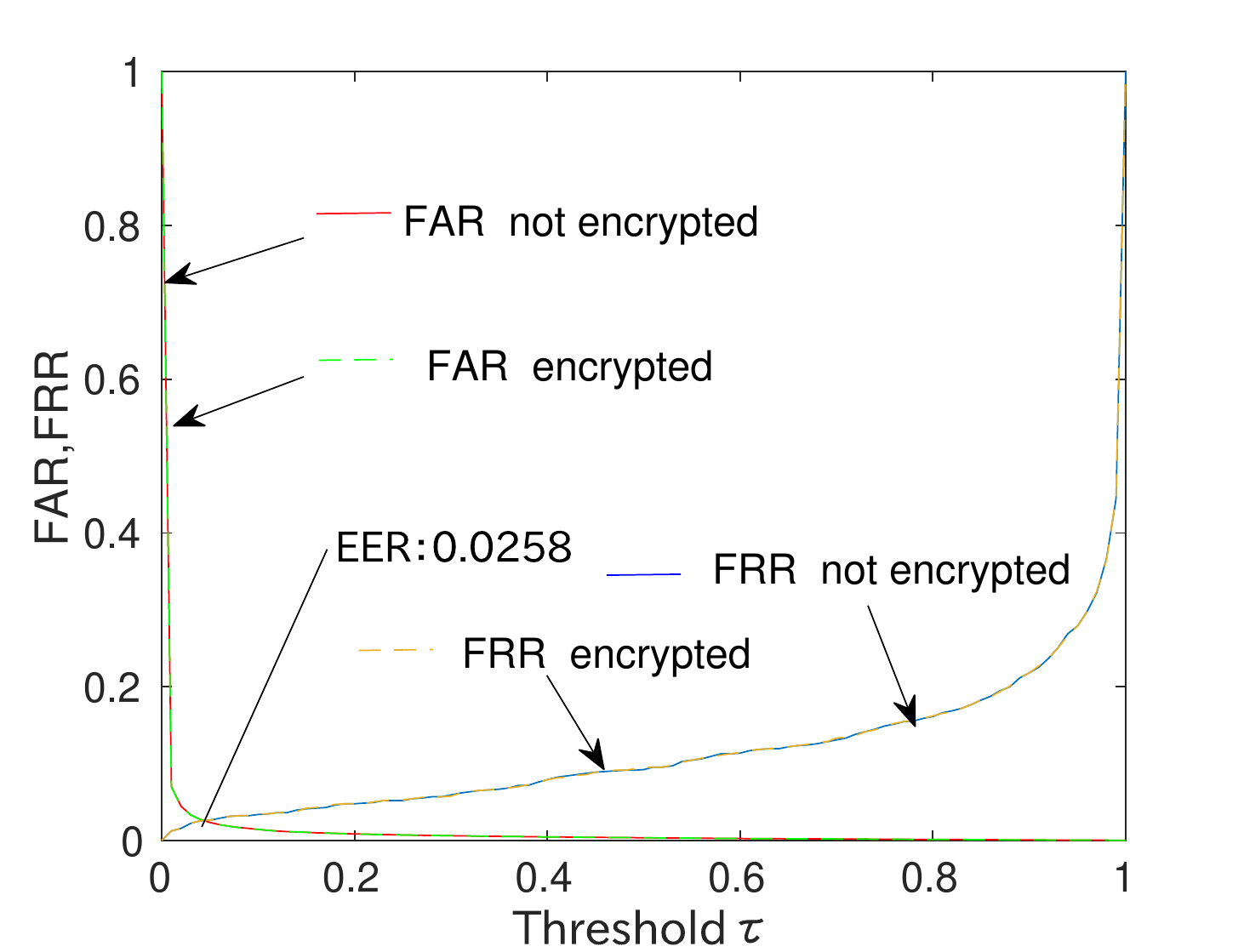}\\
  (a)Linear kernel
  \label{bs1}
 \end{minipage}\\
 \begin{minipage}[b]{1\linewidth}
  \centering
  \includegraphics[keepaspectratio, width=7.5cm]
  {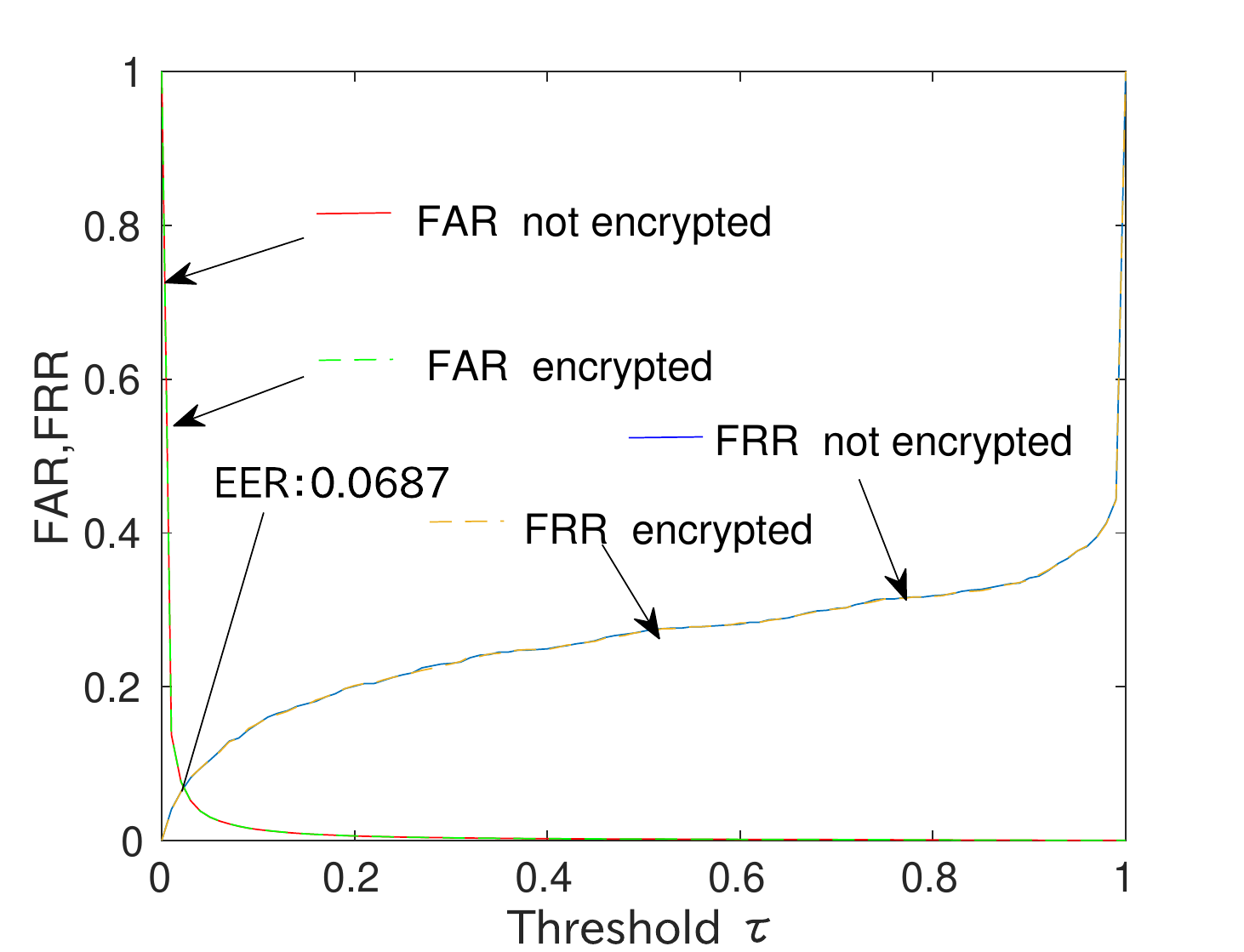}\\
  (b)RBF kernel
  \label{bs2}
 \end{minipage}
 \caption{{$\scriptstyle \mbox{FRR and FAR for SVM}\atop \scriptstyle \mbox{($\bm{B}_1=\bm{B}_2=\ldots=\bm{B}_N$, reduction ratio = 1/64)}$}}\label{result}
\end{figure}

\begin{table}[t]
\begin{center}
\caption{{$\scriptstyle \mbox{EER values for SVM with linear kernel}\atop \scriptstyle \mbox{($\bm{B}_1\neq \bm{B}_2\neq \ldots \neq \bm{B}_N$)}$}}
\label{linearkeydif}
\begin{tabular}{|c|c|c|}
\hline
\begin{tabular}{c}reduction\\ ratio\end{tabular} & \begin{tabular}{c}not\\ encrypted\end{tabular} & encrypted \\ \hline
1/20 & 0.0223 & 0.000744 \\ \hline
1/40 & 0.0247 & 0.000835 \\ \hline
1/60 & 0.0271 & 0.000777 \\ \hline
1/80 & 0.0296 & 0.000779 \\ \hline
\end{tabular}
\end{center}
\end{table}

\begin{table}[t]
\begin{center}
\caption{{$\scriptstyle \mbox{EER values for SVM with RBF kernel}\atop \scriptstyle \mbox{($\bm{B}_1\neq \bm{B}_2\neq \ldots \neq \bm{B}_N$)}$}}
\label{RBFkeydif}
\begin{tabular}{|c|c|c|}
\hline
\begin{tabular}{c}reduction\\ ratio\end{tabular} & \begin{tabular}{c}not\\ encrypted\end{tabular} & encrypted \\ \hline
1/20 & 0.0504 & 0.000448 \\ \hline
1/40 & 0.0644 & 0.00112 \\ \hline
1/60 & 0.0732 & 0.00779 \\ \hline
1/80 & 0.0863 & 0.00855 \\ \hline
\end{tabular}
\end{center}
\end{table}

\vspace{2ex}
\noindent \textcolor{black}{A. $\bm{B}_1=\bm{B}_2=\ldots=\bm{B}_N$}

\textcolor{black}{The results with key condition 1 are shown in Figure \ref{result}.}
The results demonstrate that SVM classifiers with encrypted images ("encrypted" in Fig.\ref{result}) have the same performance as SVM classifiers with the original images ("not encrypted" in Fig.\ref{result}).

\red{In the experiment, 32 images of person 1 were used as query ones, and the FRR value of person 1 ($FRR_1$) under a $\tau$ value was calculated as follows. 
The number of images $r_1$, which were rejected as another person from Eq.(\ref{accept}), was calculated, and then the rate of the rejected images was calculated as $FRR_1 = r_1 / 32$.
Finally, the average of $FRR_i$ values over 38 people was obtained as $FRR = \sum_{i=1}^{38}(FRR_i / 38)$.}

\red{
The FAR value of person 1 ($FAR_1$) under a $\tau$ value was calculated as follows. When $37 \times 32$ images without images of person 1 were used as query ones, the number of images $s_1$, which were accepted as person 1 from Eq.(\ref{accept}), was calculated, and then the rate of the accepted images was calculated as $FAR_1 = s_1 / (37 \times 32)$. 
Finally, the average of $FAR_i$ values over 38 people was obtained as $FAR = \sum_{i=1}^{38}(FAR_i / 38)$.
}

From the results, the proposed \red{scheme was} confirmed to give no influence \red{on} the performance of SVM classifiers under key condition 1 and z-score normalization.

\vspace{2ex}
\noindent \textcolor{black}{B. $\bm{B}_1\neq \bm{B}_2\neq\ldots\neq \bm{B}_N$}

 Table\red{s} \ref{linearkeydif} and \ref{RBFkeydif} show results under using key condition 2.
In this condition, it is expected that a query is authenticated only when the query meets two requirements, i.e.\red{,} the same key and the same person, although only the same person \red{is} required under key condition 1.
Therefore, the performance of using key condition 2 is different from \red{that with} plain \red{images}, so the EER values of \red{"}encrypted\red{"} in Tables \ref{linearkeydif} and \ref{RBFkeydif} were smaller than those of \red{"}not encrypted\red{"} due to the strict requirements.
In other words, the EER values under key condition 2 outperformed those under key condition 1.

In addition, the use of key condition 2 is expected to enhance the robustness of against spoofing attacks as demonstrated in \cite{maekawa2019}.

\begin{table}[t]
\begin{center}
\caption{{$\scriptstyle \mbox{EER values for random forests }\atop \scriptstyle \mbox{($\bm{B}_1=\bm{B}_2=\ldots=\bm{B}_N$)}$}}
\label{resultRF}
\begin{tabular}{|c|c|c|}
\hline
\begin{tabular}{c}reduction\\ ratio\end{tabular} & \begin{tabular}{c}not\\ encrypted\end{tabular} & encrypted \\ \hline
1/20 & 0.104 & 0.106 \\ \hline
1/40 & 0.113 & 0.115 \\ \hline
1/60 & 0.121 & 0.121 \\ \hline
1/80 & 0.124 & 0.125 \\ \hline
\end{tabular}
\end{center}
\end{table}

\subsection{Experimental results with random forests}
 We applied the proposed method to random forests.
 Table \ref{resultRF} shows results in the case of using random forests under key condition 1.
The results demonstrate that random forests with encrypted images \red{performed} almost \red{the} same \red{as} random forests with the original images (not encrypted).
Therefore, the proposed method is shown to be applicable to random forests under key condition 1. \textcolor{black}{We also confirmed that the trendy of the random forests with key condition 2, which the EER values are smaller than those of non-encrypted images in Tables 4 and 5, is similar to that of the SVM with key condition 2.}

\section{Conclusion}
\label{sec:con}
 In this paper, we proposed a privacy-preserving machine learning scheme with EtC images.
 A novel property of EtC images was considered, and all encryption steps for generating EtC images were shown to be able to preserve the inner products between vectors under the use of z-score normalization. 
 The property allows us to apply EtC images to not only machine learning algorithms based on Euclidean distances or inner products, but also kernel trick, without any degradation in classification performance. 
 A number of facial recognition experiments using SVM and random forests were carried out to confirm the effectiveness of the proposed scheme.

\subsection*{Acknowledgements}
This work was partially supported by Grant-in-Aid for Scientific
Research(B), No.17H03267, from the Japan Society
for the Promotion Science.

\bibliographystyle{ieicetr}
\bibliography{refs}

\profile[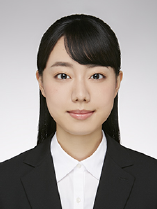]{Ayana Kawamura}{
received her B.Eng. degree
from Tokyo Metropolitan University, Japan in
2018. Since 2018, she has been a Master course
student at Tokyo Metropolitan University.
She received SIS Young Researcher Award in 2018, EMM Best student paper award in 2019, and International Workshop on Advanced Image Technology（IWAIT）BEST PAPER AWARD in 2020.
Her research interests are in the area of image processing.}


\profile[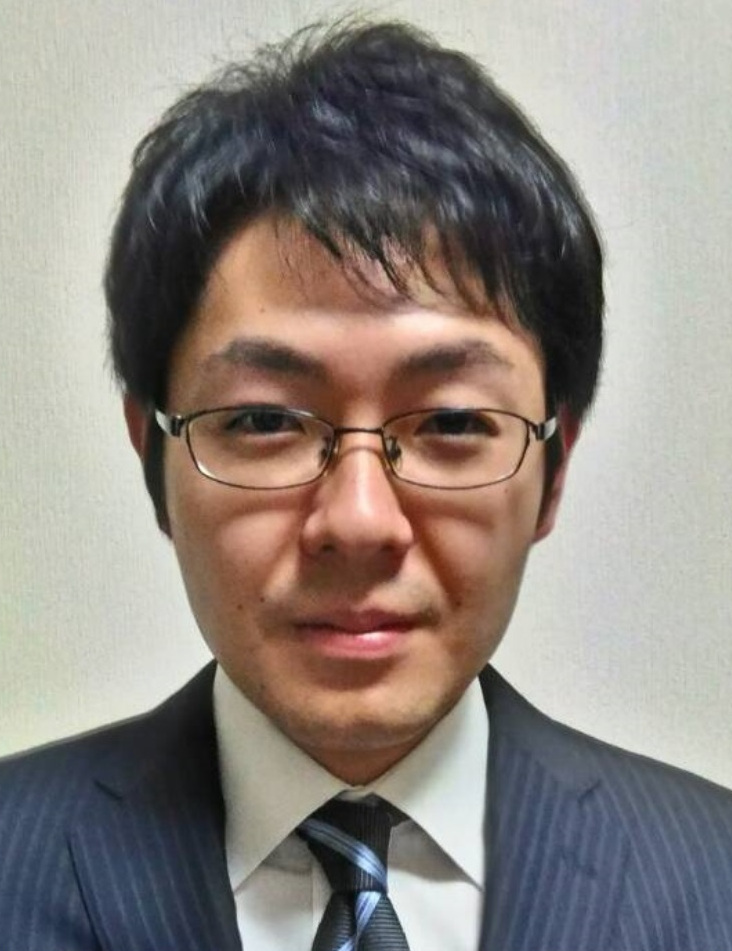]{Yuma Kinoshita}{received his B.Eng. and M.Eng. degrees from Tokyo Metropolitan University, Japan, in 2016 and 2018, respectively. From 2018, he
has been a Ph.D. student at Tokyo Metropolitan University. He received IEEE ISPACS Best Paper
Award in 2016, IEEE Signal Processing Society Japan Student Conference Paper Award in 2018, and
IEEE Signal Processing Society Tokyo Joint Chapter Student Award in 2018, respectively. His research interests are in the area of image processing. He is a student member of IEEE and IEICE.}

\profile[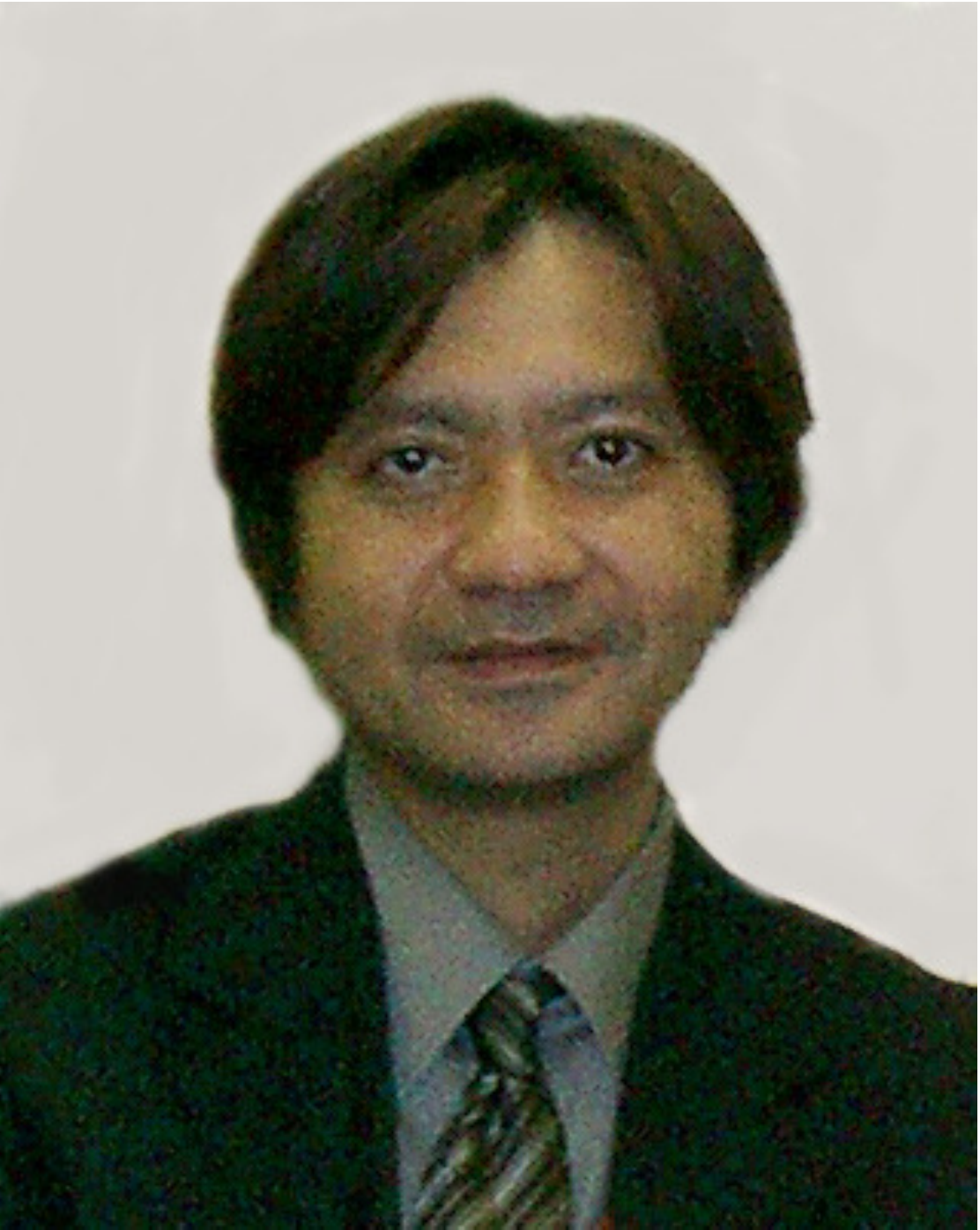]{Takayuki Nakachi}{received the Ph.D. degree in electrical engineering from Keio University, Tokyo, Japan, in 1997. Since he joined Nippon Telegraph and Telephone Corporation (NTT) in 1997, he has been engaged in research on super-high-definition image/video coding, media transport technologies. From 2006 to 2007, he was a visiting scientist at Stanford University. He also actively participates in MPEG international standardization. His current research interests include communication science, information theory and signal processing.
He received the 26th TELECOM System Technology Award, the 6th Paper Award of Journal of Signal Processing and the Best Paper Award of IEEE ISPACS2015. Dr. Nakachi is a member of the Institute of Electrical and Electronics Engineers the Institute of Electronics (IEEE) and the Information and Communication Engineers (IEICE) of Japan.}

\profile[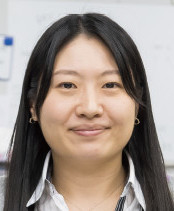]{Sayaka Shiota}{received the B.E., M.E.and Ph.D. degrees in intelligence and computer science, Engineering and engineering simulation from Nagoya Institute of Technology, Nagoya, Japan in 2007, 2009 and 2012, respectively. From February 2013 to March 2014, she had worked at the Institute of Statistical Mathematics as a project assistant professor. In April of 2014, she joined Tokyo Metropolitan University as an Assistant Professor. Her research interests include statistical speech recognition and speaker verification. She is a member of Acoustical Society of Japan (ASJ), IPSJ, IEICE, APSIPA, and IEEE.}

\profile[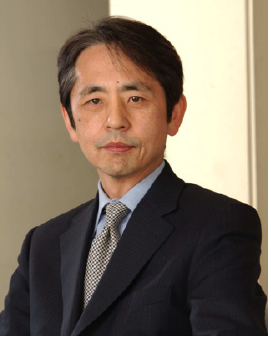]{Hitoshi Kiya}{received his B.E and M.E. degrees from Nagaoka University of Technology, in 1980 and 1982 respectively, and his Dr. Eng. degree from Tokyo Metropolitan University in 1987. In 1982, he joined Tokyo Metropolitan University, where he became a Full Professor in 2000. From 1995 to 1996, he attended the University of Sydney, Australia as a Visiting Fellow. He is a Fellow of IEEE, IEICE and ITE. He currently serves as President-Elect of APSIPA, and he served as Inaugural Vice President (Technical Activities) of APSIPA from 2009 to 2013, and as Regional Director-at-Large for Region 10 of the IEEE Signal Processing Society from 2016 to 2017. He was also President of the IEICE Engineering Sciences Society from 2011 to 2012, and he served there as a Vice President and Editor-in-Chief for IEICE Society Magazine and Society Publications. He was Editorial Board Member of eight journals, including IEEE Trans. on Signal Processing, Image Processing, and Information Forensics and Security, Chair of two technical committees and Member of nine technical committees including APSIPA Image, Video, and Multimedia Technical Committee (TC), and IEEE Information Forensics and Security TC. He has organized a lot of international conferences, in such roles as TPC Chair of IEEE ICASSP 2012 and as General Co-Chair of IEEE ISCAS 2019. He has received numerous awards, including six best paper awards.}
\end{document}